\documentclass[12pt]{article}%
\usepackage{graphicx}
\usepackage[english]{babel}
 
\begin{document}

\begin{center}{\large\bf Why Finite Mathematics Is The Most Fundamental and Ultimate Quantum Theory
Will Be Based on Finite Mathematics} \end{center}

\vskip 1em \begin{center} {\large Felix M. Lev} \end{center}
\vskip 1em \begin{center} {\it Artwork Conversion Software Inc.,
1201 Morningside Drive, Manhattan Beach, CA 90266, USA
(Email:  felixlev314@gmail.com)} \end{center}

\begin{abstract}
Classical mathematics (involving such notions as infinitely small/large and continuity) 
is usually treated as fundamental while finite mathematics is treated as inferior which is used only
in special applications. We first argue that 
the situation is the opposite: classical mathematics is only a degenerate special case of finite one
and finite mathematics is more pertinent for describing nature than standard one. Then 
we describe results of a quantum theory based on finite mathematics. Implications for foundation of
mathematics are discussed.
\end{abstract}

\begin{flushleft} Keywords: classical mathematics, finite mathematics, quantum theory\end{flushleft}

\section{Motivation}
\label{intro}

A belief of the overwhelming majority of scientists is that classical mathematics
(involving the notions of infinitely small/large and continuity) is fundamental while finite mathematics 
is something inferior what is used only in special applications. This belief is based on the fact that
the history of mankind undoubtedly shows that classical mathematics has demonstrated its
power in many areas of science. Nevertheless a problem arises whether classical mathematics is 
pertinent for constructing the ultimate quantum theory.
At present, in spite of efforts of thousands of highly qualified physicists 
and mathematicians to construct such a theory on the basis of classical mathematics, this problem has not
been solved. 

Historically the notions of infinitely small/large, continuity etc. have arisen from a belief based on everyday experience that any macroscopic 
object can be divided into arbitrarily large number of arbitrarily small parts. 
Classical physics is
based on classical mathematics developed mainly when people did not know about
existence of elementary particles. However, from the point of view of the present knowledge those notions look problematic. 

For example, a glass of water contains approximately $10^{25}$ molecules.
We can divide this water by ten, million, etc. but when we reach the level of atoms and elementary particles
the division operation loses its meaning and
we cannot obtain arbitrarily small parts. So, {\it any description of macroscopic phenomena using continuity and differentiability can be only approximate}. 
In nature there are no continuous 
curves and surfaces. For example, if we draw a line on a sheet paper and look at this line by
a microscope then we will see that the line is strongly discontinuous because it consists of
atoms.

The official birth of quantum theory is 1925, and even the word "quantum" reflects a belief that
nature is discrete. The founders of this theory were highly educated physicists
but they knew only classical mathematics because even now mathematical education at physics departments
does not involve discrete and finite mathematics. In view of the above remarks it is reasonable to think that in 
quantum theory classical mathematics
might be used for solving special problems but ultimate quantum theory should not be based on classical mathematics.

Classical mathematics is not in the spirit of the philosophy of quantum theory and the Viennese 
school of  logical positivism that {\it "A proposition is only cognitively meaningful if it can
be definitively and conclusively determined to be either true or false"}. For example, it cannot be determined whether the statement that $a + b = b + a$ for all natural numbers $a$ and $b$ is true or false.

Another example follows. Let us pose a problem whether 10+20 equals
30. Then we should describe an experiment which will solve this problem. Any
computer can operate only with a finite number of bits and can perform calculations
only modulo some number $p$. Say $p = 40$, then the experiment will confirm that
10+20=30 while if $p = 25$ then we will get that 10+20=5.
So the statements that 10+20=30 and even that $2 \cdot 2 = 4$ are ambiguous
because they do not contain {\it explicit} information on how they should be verified. On the other
hand, the statements
$$10 + 20 = 30\, (mod\, 40),\,\, 10 + 20 = 5\, (mod\, 25), \,\,
2 \cdot 2 = 4\, (mod\, 5),\,\, 2 \cdot 2 = 2\, (mod\, 2)$$
are well defined because they do contain such an information.
So only operations modulo some number are well defined. This example shows that 
classical mathematical is based on the implicit assumption that in principle one can have
any desired amount of resources and, in particular, one can work with computers
having as many bits as desired. 

The opinion that classical mathematics is fundamental is often based on the fact
that it contains more numbers than finite mathematics. Let us consider this problem in greater details. 

Classical mathematics starts from natural numbers and the famous Kronecker's expression is: 
{\it "God made the natural numbers, all else is the work of man"}. 
However here only addition and multiplication are always possible. In order to make addition invertible we introduce negative
integers. They do not have a direct physical meaning (e.g. the phrases "I have -2 apples" or "this computer
has -100 bits of memory" are meaningless) and their only goal is to get the ring of integers $Z$. 

However, if instead of all natural numbers we consider only a set $R_p$ of 
$p$ numbers 0, 1, 2, ... $p-1$ where addition and multiplication are defined
as usual but modulo $p$ then we get a ring without adding new elements. 
If, for example, $p$ is odd then 
one can consider $R_p$ as a set of elements $\{0,\pm i\}$ $(i=1,...(p-1)/2)$.
If elements of $Z$ are depicted as integer points on the $x$ axis of the $xy$ plane then 
it is natural to depict the elements of $R_p$ as points of the circumference in Fig. 1.
\begin{figure}[!ht]
\centerline{\scalebox{0.3}{\includegraphics{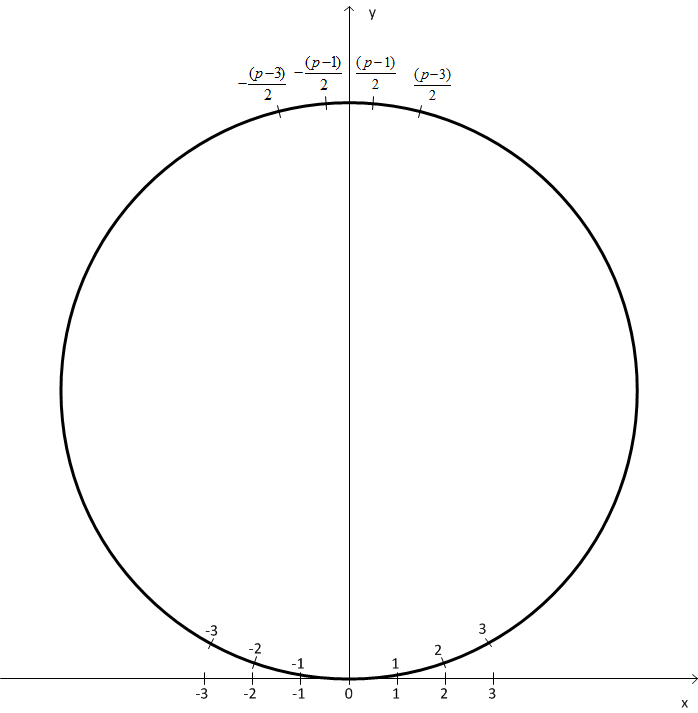}}}
\caption{
  Relation between $R_p$ and $Z$
}
\label{Fig.1}
\end{figure}

Let $f$ be a function from $R_p$ to $Z$ such that
$f(a)$ has the same notation in $Z$ as $a$ in $R_p$. Then for elements $a\in R_p$ such that $|f(a)|\ll p$, 
addition, subtraction and multiplication are the same as in $Z$. In other words, for
such elements we do not feel the existence of $p$. Indeed, for elements $a_j\in R_p$ ($j=1,2$)
such that $|f(a_j)|<[(p-1)/2]^{1/2}$ we have that $f(a_1\pm a_2)=f(a_1)\pm f(a_2)$ and $f(a_1a_2)=f(a_1)f(a_2)$
which shows that $f$ is a local isomorphism of some vicinities of zero in $R_p$ and $Z$.  

As explained in textbooks, both $R_p$ and $Z$ are cyclic groups with respect to addition. However, 
$R_p$ has a higher symmetry because, in contrast to $Z$, $R_p$ has a property which we call {\it strong cyclicity}:
for any fixed $a\in R_p$ any element of $R_p$ can be obtained from $a$ by successively adding 1.
As noted below, in quantum physics the presence or absence of strong cyclicity plays an important role. 

When $p$ increases, the bigger and bigger part of $R_p$
becomes the same as $Z$. Hence $Z$ can be treated as a degenerate case of 
$R_p$ in the formal limit $p\to\infty$ because in this limit operations modulo $p$ disappear
and strong cyclicity is broken. {\it Therefore, at the level of rings standard mathematics is a degenerate case of finite one when formally $p\to\infty$}.

The transition from $R_p$ to $Z$ is similar to the procedure, which in group theory is called contraction.
This notion is used when the Lie algebra of a group with a lower symmetry 
can be treated as a formal limit of the Lie algebra of a group with a higher symmetry when some parameter goes to zero or infinity. Known examples are the contraction from the de Sitter
to the Poincare group and from the Poincare to the Galilei group. 

The above construction has a well-known historical analogy. For many years people believed
that the Earth was flat and infinite, and only after a long period of time they realized that
it was finite and curved. It is difficult to notice the curvature when we deal only with
distances much less than the radius of the curvature. Analogously one might think that
the set of numbers describing physics in our Universe has a "curvature" defined by a very
 large number $p$ but we do not notice it when we deal only with numbers much less than $p$. 

One might argue that introducing a new fundamental constant $p$ is not justified.
However, history of physics tells us that more general theories arise when a parameter,
which in the old theory was treated as infinitely small or infinitely large, becomes finite. 
For example, nonrelativistic physics is the degenerate case of relativistic one in the formal limit $c\to\infty$ 
and classical physics is the degenerate case of quantum one
in the formal limit $\hbar\to 0$.  Therefore, it is natural to think that
in quantum physics the quantity $p$ should be not infinitely large but finite.

The above discussion shows that when we take the formal limit $p\to\infty$ then we obtain a less
general theory, not a more general one. Therefore the fact that classical mathematics contains more
numbers than finite one does not indicate that classical mathematics is more fundamental.

From mathematical point of view standard quantum theory can be treated as a theory of
representations of special real Lie algebras in complex Hilbert spaces. In Refs. \cite{lev4,PRD} and
other publications we have
proposed an approach called FQT (Finite Quantum Theory) when Lie algebras and representation spaces are over a finite field or ring with characteristic $p$. It has been shown that in the formal limit
$p\to\infty$ FQT recovers predictions of standard continuous theory.
Therefore classical mathematics describes many experiments with a high accuracy as a
consequence of the fact that the number p is very large.

In Sec. \ref{S1} we consider several problems where it is important that the number $p$
is finite and not infinite and Sec. \ref{conclusion} is a discussion.

\section{Main results of FQT}
\label{S1}

\subsection{Particles and antiparticles in FQT}

A known fact of particle physics is that a particle and its antiparticle have equal masses.
The explanation of this fact in quantum field theory (QFT) follows. 
Irreducible representations (IRs) of the Poincare and anti-de Sitter (AdS) algebras by Hermitian operators
used for describing elementary particles have the property that for each IR the Hamiltonian is either positive definite or negative definite. In the first case, the energy has the spectrum in the range $[m_1,\infty)$, while in the second case it has the spectrum in the range $(-\infty,-m_2]$ ($m_1,m_2\geq 0$). The quantities $m_1$ and $m_2$ are called the masses of the
particle and its antiparticle, respectively.

Hence a particle and its antiparticle are described by different IRs and the equality $m_1=m_2$ is
problematic. In QFT this equality follows from the assumption that a particle
and its antiparticle can be described by a local field satisfying a covariant equation (e.g. the Dirac equation). However, a problem arises whether this equality remains if locality is only approximate.
In addition, the meaning of locality is not quite clear because local fields do not
have probabilistic interpretation. 

By analogy with standard quantum theory, it is natural to define the elementary particle in FQT as
a system described by an IR of a Lie algebra over a finite field or ring with characteristic $p$. Representations of Lie algebras in spaces with nonzero characteristic are called 
modular representations and there exists a well developed theory of such representations. 

To illustrate the difference between the treatment of particles and antiparticles in standard theory
and in FQT we first note how IRs for the AdS algebra are described
in standard theory. In the case of a particle one starts from the "rest" state $e_1$  where
energy equals $m_1$. When the representation operators act on $e_1$ one obtains
states with higher and higher energies and the energy spectrum is in the range $[m_1,\infty )$.
On the other hand, for describing antiparticles one starts from the state $e_2$ where
energy equals $-m_2$. When the representation operators act on $e_2$ one obtains
states with lower and lower energies and the energy spectrum is in the range $( -\infty,-m_2]$.

However, in FQT one can prove \cite{monograph} that: {\it The IR with the cyclic
vector $e_1$ and the IR with the cyclic vector $e_2$ are the same and $m_1=m_2$.}
The explanation follows. When the representation operators act on $e_1$ and increase the energy,
then, since the values of the energy now belong not to $Z$ but to $R_p$, we are moving not
in the positive direction of the $x$ axis but along the circumference in Fig. 1. Then, as a consequence of
strong cyclicity, sooner or 
later we will arrive to states where the energy is "negative" (i.e. in the range $[-(p-1)/2,-1]$) 
and finally we will arrive to 
the state where the energy equals $-m_1$. From the point of view of physics this means that one
modular IR describes a particle and its antiparticle simultaneously. 

As shown in Ref. \cite{JPA}, in standard theory a particle and its antiparticle are described
by the same IR in a special case when the theory is based on de Sitter (dS) symmetry. However, in
FQT this is true for any symmetry as a consequence of strong cyclicity.

As a consequence, in FQT a particle and its antiparticle automatically have the same masses.
Moreover, while in standard theory the existence of antiparticles depends on additional assumptions, in FQT it is inevitable. Therefore, {\it the very existence of antiparticles is a strong indication that nature 
is described by a finite field or ring rather than by complex numbers.}

Since a particle and its antiparticle belong to the same IR, transitions $particle\leftrightarrow antiparticle$
are not prohibited. This means that the very notions of a particle and its antiparticle are only
approximate and the conservation of the electric charge and the baryon and lepton quantum
number is approximate too. At present the corresponding conservation laws work with a high
accuracy because $p$ is very large (see below) and the dS energies are much less than $p$.
However, a reasonable possibility is that at early stages of the Universe $p$ was not as large as now.
This might explain the problem known as the baryon asymmetry of the Universe (see Ref. 
\cite{monograph} for a detailed discussion).

\subsection{Vacuum energy in FQT}

The vacuum energy problem is discussed in practically
every textbook on quantum field theory. Its essence is as follows. 
After quantization the energy operator can be written in the form $E={\tilde E}+E_{vac}$
where ${\tilde E}$ has the normal form (when creation operators always precede annihilation
ones) and $E_{vac}$ is a constant. In the vacuum state, when particles are absent, the energy equals
$E_{vac}$ and therefore $E_{vac}$ should be zero. However, the actual calculation
shows that it is infinite. This is an indication
that standard theory has consistency problems.

In FQT for calculating the vacuum energy
one should break all possible states into two parts which can be treated as physical
and nonphysical ones, respectively. As explained in Ref. \cite{monograph}, this can be achieved
only for fermions with the mass $m$ and spin $s$ such that in dS units $f(m)$ and $f(s)$ are odd. Then 
a direct calculation (see Ref. \cite{monograph}) gives that 
\begin{equation} 
E_{vac}=(m-3)(s-1)(s+1)^2(s+3)/96
\end{equation}

Our conclusion is that while in standard theory the vacuum energy is
infinite, in FQT it is not only finite (in finite mathematics it
cannot be infinite) but is exactly zero if $s=1$ (i.e. $s=1/2$ in the usual units). Note 
that if $p$ is treated 
only as a regulator
then the vacuum energy would be a quantity which depends on $p$ and becomes infinitely
large in the formal limit $p\to\infty$. However, since the rules of arithmetic in finite mathematics are
different from those for complex numbers the vacuum energy is exactly zero as it should be.

Existence of infinities is one
of the main problems in constructing standard quantum theory. However, in FQT infinities 
cannot exist in principle and the above example is a clear demonstration of
this fact.

\subsection{Cosmological constant and gravity}
\label{gravity}

The philosophy of general relativity (GR) is that the curvature of space-time is defined by 
matter in that space-time. Therefore empty space-time should be flat, i.e. the cosmological
constant (CC) $\Lambda$ should be zero. This was one of the subjects of the debates 
between Einstein and
de Sitter. However, the phenomenon of the cosmological acceleration discovered in 1998
is interpreted such that 
$\Lambda >0$ with the accuracy better than 5\%. To reconcile this fact with the requirement
$\Lambda=0$ the term with $\Lambda$ in the Einstein equations is moved from the
l.h.s. to the r.h.s. and is interpreted not as the curvature of empty space-time but as
dark energy.

In QFT one starts from the choice of the space-time background. The background has
the symmetry group and the operators characterizing the system under consideration should
satisfy the commutation relation of the Lie algebra for this group. By analogy with the philosophy of GR
it is believed that the choice of the Minkowski background is more physical than the choice
of the dS one and that the goal of quantum gravity is to explain the 
value of $\Lambda$. The existing quantum theory of gravity contains strong
divergencies and with a reasonable cutoff the theory gives for $\Lambda$ a value exceeding
the experimental one by 122 orders of magnitude. This is called the CC problem.

However, the physical meaning of the curvature is to describe the motions of bodies. Therefore
the curvature of empty space-time is only a mathematical notion which does not have a
physical meaning. As argued in Ref. \cite{PRD}, the 
approach should be opposite to standard one. Every quantum system is described by a set of
operators which somehow commute with each other and the rules of their commutation define
the symmetry algebra. Therefore in quantum theory one should start not from the space-time
background, which is the classical notion, but from the symmetry algebra. From this
point of view the dS symmetry is more preferable than the Poincare one because
the Poincare algebra is less symmetric than the dS one and can be obtained from
the latter by contraction. 

In Poincare invariant theory the mass operator of the free two-body system depends
only on relative momenta but not on relative distances. As a consequence, in semiclassical
approximation the relative acceleration of the bodies is zero. Consider now what happens 
when one starts from the dS algebra.

The problem contains the dS radius $R$ (the radius of the Universe) because
instead of working with dimensionless dS angular momenta $M^{4\mu}$ $(\mu=0,1,2,3)$ which are fundamental we wish to work with the Poincare four-momenta $P^{\mu}=M^{4\mu}/R$.
The problem why $R$ is as is does not arise since the answer is: because we wish to measure
distances in meters. Then, as shown in Ref. \cite{JPA},
in semiclassical approximation one recovers the same result for
the relative acceleration as in GR for the dS background {\it if we denote} $\Lambda=3/R^2$.

However, although the result is the same its interpretation fully differs from that in GR.
This result has been obtained without using space-time background and Riemannian geometry
(metric tensor, connection etc.) but only in the framework of standard quantum mechanical 
calculation. In that case $\Lambda$ has nothing to do with the curvature of the background space-time
and has no relation to the value of the gravitational constant $G$. Therefore
the problem of explaining the value of $\Lambda$ does not arise,
the CC problem does not exist and
for explaining the cosmological acceleration there is no need to involve empty space-time and
dark energy. 

In dS theory the spectrum of the {\it free} two-body mass operator is not bounded below by
$m_1+m_2$, where $m_1$ and $m_2$ are the masses of the particles, but contains
values less than $m_1+m_2$. There is no law prohibiting that in semiclassical approximation
the mean value of the {\it free} mass operator contains the term $-Gm_1m_2/r$ with possible 
corrections. Here $r$ is the
relative distance and $G$ is not a quantity taken from the outside but a value which should
be calculated.

As shown in Ref. \cite{monograph}, for macroscopic bodies standard distance operator
is semiclassical only at cosmological distances. Therefore this operator should be
modified. We propose a new distance operator which satisfies all the required properties.
Then a detailed calculation \cite{monograph} gives that in quantum theory based on representations of
the dS algebra the mean value of the Hamiltonian for a system of two free nonrelativistic bodies is
given by
\begin{equation}
H = H_0 - \frac{Cm_1m_2}{(m_1+m_2)r}(\frac{1}{\delta_1}+\frac{1}{\delta_2})
\label{dS}
\end{equation}
where $H_0$ is the free Hamiltonian, $C$ is a constant and $\delta_i$ ($i=1,2)$ 
is the width of the dS momentum distribution for body $i$. 
The last term in Eq. (\ref{dS}) represents the dS correction to standard expression. 
We see that the correction disappears if the width of the 
momentum distribution for each body becomes very large. In standard theory
there is no limitation on the width and this correction is negligible.

However, in FQT the width cannot be arbitrarily large. 
Suppose that a macroscopic body 
consists of $N$ components and  $\delta_j$ ($j=1,2,...N$) is the width of the momentum distribution
of the $j$th component with the mass $m_j$. Then, as shown in Ref. \cite{monograph}, a
necessary condition for a wave 
function to have a probabilistic interpretation is 
\begin{equation}
R\sum_{j=1}^N \delta_j lnw_j \ll lnp
\label{70}
\end{equation} 
where $w_j=4R^2m_j^2$. 
This condition shows that the greater the number of components is,
the stronger is the restriction on the width of the momentum distribution for each component.
This  is a crucial difference between standard theory and FQT. A naive explanation 
is that if $p$ is
finite, the same set of numbers which was used for describing one body is now shared between $N$ bodies.
In other words, if in standard theory each body in the free $N$-body system does not feel the presence of
other bodies, in FQT this is not the case. 

The existing theory does not make it possible to reliably calculate the width of the total momentum
distribution for a macroscopic body and at best only a qualitative estimation of this quantity can be
given. Equation (\ref{70}) indicates that the quantities $\delta_i$ in Eq. (\ref{dS}) are inversely 
proportional to the corresponding masses $m_i$ in agreement with the Newton law.
A detailed discussion in Ref. \cite{monograph} shows that  with reasonable assumptions
the result given by 
Eq. (\ref{dS}) can be written in the form $H=H_0-Gm_1m_2/r$ where $G=const\, R/(m_0lnp)$
and $m_0$ is the nucleon mass. We also discuss whether relativistic corrections to
the Newton gravity law are compatible with GR.

We conclude that in FQT the phenomenon of gravity can be treated not as 
an interaction but simply
as a consequence of the fact that $p$ is finite. If we assume that $const$ is of the order of unity and take for $R$ a reasonable value $R=10^{26}m$ then the comparison with
the experimental value of $G$ gives that $lnp$ is of the order of $10^{80}$. Therefore $p$ is a huge
number of the order of $exp(10^{80})$.

\subsection{Ring or field?}

In standard quantum theory states are described by elements of Hilbert spaces. Such spaces
are linear spaces over a field of complex numbers. A field is a set with four operations: addition,
subtraction, multiplication and division. 
In the literature there have been also discussed approaches where quantum theory is based on the body of quaternions,  $p$-adic fields or adelic rings built on the field of rational numbers (see e.g. Ref. \cite{Adler} and references therein). In the cellular automation interpretation
of quantum theory proposed by 't Hooft (see Ref. \cite{tHooft} and references therein) the observables can be only integers and the evolution of states with such observables is described by standard mathematics.
In all those approaches  a problem remains whether or not it is possible to construct quantum theory without infinities. 

In FQT linear spaces can be over either finite rings or finite
fields. A ring is a set with three operations: addition, subtraction and multiplication. Known
facts from standard algebra are that invariance of dimension, basis and linear independence are well defined
only in spaces over a field or body. In addition, existence of division is often convenient for
calculations.

At the same time, as argued in Sec. \ref{intro}, in quantum theory division is not fundamental.
History of physics tells us that it is desirable to construct physical theories with the least required
notions. Therefore a problem arises whether ultimate quantum theory can be constructed
without using division at all. For the first time this possibility has been discussed in Ref.
\cite{Saniga}.

As shown in Ref. \cite{monograph}, modular IRs describing massive and massless particles
can be constructed only in spaces over a field. However, in Ref. \cite{DiracS} titled 
"A Remarkable Representation of the 3 + 2 de Sitter group" Dirac discovered
the existence of a new type of particles - Dirac singletons. In Standard Model only massless
particles are fundamental but, as shown in Ref. \cite{FF}, massless particles can be constructed
from singletons. This poses a problem whether singletons are the only true fundamental
particles.

As discussed in Ref. \cite{monograph}, in FQT Dirac singletons are even more remarkable
than in standard theory and the singleton physics can be constructed only over a ring.
This poses a problem whether the ultimate quantum theory will be constructed over a
ring, not a field.

\section{Discussion and conclusion}
\label{conclusion}

In Sec. \ref{intro} we argue that classical mathematics is a degenerate case of finite one in the formal limit
$p\to\infty$ and that ultimate quantum theory will be based on finite mathematics. The results described in Sec. \ref{S1} can be treated as arguments in favor of this statement. 

The estimation of the gravitational constant shows that at present physics in our Universe is described
by finite mathematics such that $p$ is a huge number of the order of $exp(10^{80})$. Nevertheless, gravity is
a manifestation of the fact that $p$ is finite and not infinite. The matter is that the gravitational constant
depends on $p$ as $1/lnp$. Therefore in the formal limit $p\to\infty$ this constant disappears.
The fact that $p$ is a huge number explains why in many cases classical mathematics describes natural phenomena with a very high accuracy. 
At the same time, the above discussion shows that the explanation of several phenomena can be given only
in the theory where $p$ is finite.

Although classical mathematics is a degenerate case of finite one, a problem arises
whether classical mathematics can be substantiated as an abstract science.
It is well-known that, in spite of great efforts of many great mathematicians, the problem of foundation of
classical mathematics has not been solved. For example, G\"{o}del's incompleteness theorems state that no system of axioms can ensure that all facts about natural numbers can be proven and the system of axioms in classical mathematics cannot demonstrate its own consistency.

The philosophy of Cantor, Fraenkel, G\"{o}del, Hilbert, Kronecker, Russell, Zermelo and 
other great mathematicians was based on macroscopic experience in which the 
notions of infinitely small, infinitely large, continuity and standard division are natural. 
However, as noted above, those notions contradict the existence of elementary particles and are not natural
in quantum theory. The illusion of continuity arises when one neglects the discrete structure of matter.

However, since classical mathematics is a special degenerate case of finite one,
foundational problems in this mathematics do not have a fundamental role and 
classical mathematics can be treated only as a 
technique which in many cases (but not all of them) describes reality with a
high accuracy.

{\bf Acknowledgments:} I am grateful to participants of the Fq12 Conference, Volodya Netchitailo,
Metod Saniga, Teodor Shtilkind and the (anonymous) referee of this paper for important remarks and discussions.


\begin{thebibliography}{99}
\bibitem{lev4} F. Lev, J. Math. Phys., {\bf 30}, 1985 (1989); J. Math. Phys. {\bf 34}, 490 (1993);
Finite Fields and Their Applications {\bf 12}, 336 (2006).
\bibitem{PRD} F. Lev, Phys. Rev. {\bf D85} 065003 (2012).
\bibitem{monograph} F. Lev, arxiv:1104.4647.
\bibitem{JPA} F.M. Lev, J. Phys. A: Mathematical and Theoretical {\bf 37}, 3287 (2004).
\bibitem{Adler} S.L. Adler, {\it Quaternionic Quantum Mechanics and Quantum Fields}. Oxford University Press: 
Oxford England (1995); B. Dragovich, A. Yu. Khrennikov, S. V. Kozyrev, and I. V. Volovich,
p-Adic Numbers, Ultrametric Analysis and Applications {\bf 1}, 1 (2009);
B. Dragovich, arXiv:0707.3876.
\bibitem{tHooft} G. 't Hooft, arxiv:1405.1548.
\bibitem{Saniga}  M. Saniga and M. Planat, J. Mod. Phys. {\bf B20}, 1885 (2006).
\bibitem{DiracS}  P.A.M. Dirac, J. Math. Phys. {\bf 4}, 901 (1963).
\bibitem{FF} M. Flato and C. Fronsdal, Lett. Math. Phys. {\bf 2}, 421 (1978).

\end{thebibliography}
\end{document}